\documentclass{birkmult}
\usepackage{euscript,amssymb,amsmath,mathdots}

\newcommand{\be}[1]{\begin{equation}\label{#1}}
\newcommand{\ee}{\end{equation}}
\newcommand{\ba}[1]{\begin{eqnarray}\label{#1}}
\newcommand{\ea}{\end{eqnarray}}
\newcommand{\rf}[1]{(\ref{#1})}
\newcommand{\nn}{\nonumber}

\def\RR{\mathbb{R}}

\def\CC{\mathbb{C}}

\newcommand{\bb}{\mathbf{b}}

\newcommand{\bl}{\mathbf{l}}
\newcommand{\bu}{\mathbf{u}}
\newcommand{\bv}{\mathbf{v}}
\newcommand{\bw}{\mathbf{w}}
\newcommand{\bA}{\mathbf{A}}
\newcommand{\bB}{\mathbf{B}}

\newcommand{\bU}{\mathbf{U}}

\newcommand{\cL}{\mathcal{L}}
\newcommand{\cU}{\mathcal{U}}
\newcommand{\cV}{\mathcal{V}}

\newcommand{\fb}{\mathfrak{b}}

\newcommand{\fu}{\mathfrak{u}}
\newcommand{\fv}{\mathfrak{v}}
\newcommand{\fw}{\mathfrak{w}}

\newcommand{\fA}{\mathfrak{A}}
\newcommand{\fB}{\mathfrak{B}}
\newcommand{\fC}{\mathfrak{C}}
\newcommand{\fD}{\mathfrak{D}}

\newcommand{\fL}{\mathfrak{L}}
\newcommand{\fU}{\mathfrak{U}}
\newcommand{\fV}{\mathfrak{V}}

\begin{document}
%
%
%
%
%
%
%
%
%
\title[Multiparameter perturbation theory for non-self-adjoint operator matrices]
 {Perturbation of multiparameter \\ non-self-adjoint boundary eigenvalue \\
 problems for
  operator matrices}
\author[O.N. Kirillov]{O.N. Kirillov}

\address{%
Dynamics and Vibrations Group\\
Department of Mechanical Engineering\\
Technische Universit\"at Darmstadt\\
Hochschulstr. 1,
64289 Darmstadt,
Germany}

\email{kirillov@dyn.tu-darmstadt.de}

\thanks{This work was completed with the support of the Alexander von Humboldt Foundation and the grant DFG HA 1060/43-1}


\date{March 14, 2008}

\begin{abstract}
We consider two-point non-self-adjoint boundary eigenvalue problems for linear
matrix differential operators. The coefficient matrices in the differential
expressions and the matrix boundary conditions are assumed to depend analytically
on the complex spectral parameter $\lambda$ and on the vector of real physical
parameters ${\bf p}$. We study perturbations of semi-simple multiple eigenvalues
as well as perturbations of non-derogatory
eigenvalues under small variations of ${\bf p}$. Explicit formulae describing the
bifurcation of the eigenvalues are derived.
Application to the problem of excitation of unstable modes in rotating continua such as spherically symmetric
MHD $\alpha^2$-dynamo and circular string demonstrates
the efficiency and applicability of the theory.

\end{abstract}

\maketitle
\section{Introduction}

Non-self-adjoint boundary eigenvalue problems for matrix
differential operators describe distributed non-conservative
systems with the coupled modes and appear in structural mechanics,
fluid dynamics, magnetohydrodynamics, to name a few.

Practical needs for optimization and rational experiment planning in modern applications
allow both the differential expression and the boundary conditions to depend
analytically on the spectral parameter and smoothly on several
physical parameters (which can be scalar or distributed). According to the ideas going back to
von Neumann and Wigner \cite{NW29}, in the multiparameter operator families, eigenvalues
with various algebraic and geometric multiplicities can be generic \cite{Ar83}.
In some applications additional symmetries yield the existence of \textit{spectral meshes} \cite{GK06} in the plane \lq eigenvalue
versus parameter' containing infinite number of nodes with the multiple eigenvalues \cite{L74,YH95,VV05,Ki08}.
As it has been pointed out already by Rellich \cite{Re37} sensitivity analysis of multiple
eigenvalues is complicated by their non-differentiability as functions of several parameters.
Singularities corresponding to the multiple eigenvalues \cite{Ar83} are related to such important effects
as destabilization paradox in near-Hamiltonian and near-reversible systems \cite{Ho61,BRS02,Ki05,KS05,KS05a,KM07},
geometric phase \cite{MKS05},
reversals of the orientation of the magnetic field in MHD dynamo models \cite{SGGX06}, emission of sound by
rotating continua interacting with the friction pads \cite{Ki08} and other phenomena \cite{SM03}.

An increasing number of multiparameter non-self-adjoint boundary eigenvalue problems and
the need for simple constructive estimates of critical parameters and eigenvalues as well as for verification of
numerical codes, require development of applicable methods, allowing one to track relatively easily
and conveniently the changes in simple and multiple eigenvalues and the corresponding eigenvectors due to variation
of the differential expression and especially due to transition from one type of boundary conditions to another one
without discretization
of the original distributed problem, see e.g. \cite{P77,YH95,BRS02,KS02a,GK07,GKSS07,Ki08}.

A systematical study of bifurcation of eigenvalues of a non-self-adjoint linear operator $L_0$ due to perturbation
$L_0+\varepsilon L_1$,
where $\varepsilon$ is a small parameter, dates back to 1950s. Apparently, Krein was the first who derived a
formula for the splitting of a double eigenvalue with the Jordan block at the Hamiltonian $1:1$ resonance,
which was expressed through the generalized eigenvectors of the double eigenvalue \cite{K50}.
In 1960 Vishik and Lyusternik and in 1965 Lidskii created a perturbation theory
for nonsymmetric matrices and non-self-adjoint differential operators allowing one to find the perturbation coefficients of
eigenvalues and eigenfunctions in an explicit form by means of the eigenelements of the unperturbed operator
\cite{VL60,Li65}. Classical monographs by Rellich \cite {Re68}, Kato \cite{K66}, and Baumg\"artel \cite{Ba84}, mostly
focusing on the self-adjoint case, contain a detailed
treatment of eigenvalue problems linearly or quadratically dependent on the spectral parameter.

Recently Kirillov and Seyranian proposed a perturbation theory of multiple eigenvalues and eigenvectors
for two-point non-self-adjoint boundary eigenvalue problems with scalar differential expression and boundary conditions,
which depend analytically on the spectral parameter and smoothly on a vector of physical parameters,
and applied it to the sensitivity analysis of distributed non-conservative problems prone to dissipation-induced instabilities \cite{KS99, KS00, KS02, Ki03, KS02a, KS04, KS05, KS05a, Ki08}. An extension to the case of intermediate boundary conditions with an application
to the problem of the onset of friction-induced oscillations in the moving beam was considered in \cite{SKH07}.
In \cite{GK06} this approach was applied to the study of MHD $\alpha^2$-dynamo model with idealistic boundary conditions.

In the following we develop this theory further and consider boundary eigenvalue problems for linear non-self-adjoint
$m$-th order $N\times N$ matrix differential operators on the
interval $[0,1]\ni x$.  The coefficient matrices in the differential
expression and the matrix boundary conditions are assumed to depend
analytically on the spectral parameter $\lambda$ and smoothly on a vector of real physical
parameters ${\bf p}$. The matrix formulation of the boundary conditions is chosen for the
convenience of its implementation in computer algebra systems
for an automatic derivation of the adjoint eigenvalue problem and perturbed eigenvalues and
eigenvectors, which is especially helpful when the order of the derivatives in the differential expression is high.
Based on the eigenelements of the unperturbed problem explicit formulae are derived describing
bifurcation of the semi-simple multiple eigenvalues (diabolical
points) as well as non-derogatory
eigenvalues (branch points, exceptional points) under small variation of the parameters
in the differential expression and in the boundary conditions.
Finally, the general technique is applied to the
investigation of the onset of oscillatory instability in rotating continua.

\section{A non-self-adjoint boundary eigenvalue problem for a matrix differential operator}

Following \cite{Na67, KS02, MM03, KS04, KS05, KS05a} we consider the boundary eigenvalue problem
\be{s1}
{\bf L}(\lambda,{\bf p})\bu=0,\quad \bU_k(\lambda, {\bf p})\bu=0, \quad k=1,\ldots, m,
\ee
where $\bu(x)\in \CC^{N}\otimes C^{(m)}[0,1]$. The differential expression ${\bf L} \bu$
of the operator is
\be{s2}
{\bf L} \bu=\sum_{j=0}^{m}\bl_j(x)\partial^{m-j}_x \bu ,
\quad \bl_j (x)\in \CC^{N\times N}\otimes C^{(m-j)}[0,1], \quad \det[\bl_0(x)] \neq 0,
\ee
and the boundary forms $\bU_k \bu$ are
\be{s3}
\bU_k \bu=\sum_{j=0}^{m-1 }\bA_{kj}
\bu^{(j)}_x(x=0)+\sum_{j=0}^{m-1} \bB_{kj} \bu^{(j)}_x(x=1),  \quad
\bA_{kj},\bB_{kj}\in \CC^{N\times N}.
\ee
Introducing the block matrix
$
\fU:=[\fA,\fB]\in \CC^{mN\times 2mN}$
and the vector
\be{s6} \fu^T:=\left(\bu^T(0),\bu^{(1)T}_x(0),\ldots,
\bu^{(m-1)T}_x(0),\bu^T(1),\bu^{(1)T}_x(1),\ldots,
\bu^{(m-1)T}_x(1)\right)\in \CC^{2mN}
\ee
the boundary conditions can be compactly rewritten as \cite{KS05, KS05a}
\be{s7}
\fU
\fu=[\fA,\fB]\fu=0,
\ee
where $\fA=\left.(\bA_{kj})\right|_{x=0}\in
\CC^{mN\times mN}$ and $\fB=\left.(\bB_{kj})\right|_{x=1} \in
\CC^{mN\times mN}$.
It is assumed that the matrices $\bl_j$, $\fA$, and $\fB$
are analytic functions of the complex spectral parameter
$\lambda$ and smooth functions of the real vector of physical parameters ${\bf p}\in \RR^n$.
For some fixed vector ${\bf p}={\bf p}_0$ the eigenvalue $\lambda_0$,
to which the eigenvector ${\bf u}_0$ corresponds, is a root of the characteristic equation obtained
after substitution of the general solution to equation ${\bf Lu}=0$ into the boundary conditions \rf{s7} \cite{Na67}.

Let us introduce a scalar product \cite{Na67}
\be{s9}
<\bu ,\bv >:=\int_0^1  \bv^* \bu dx,
\ee
where the asterisk denotes complex-conjugate transpose ($\bv^*:=\overline \bv^T$).
Taking the scalar product of ${\bf L}{\bf u}$ and a vector-function ${\bf v}$
and integrating it by parts yields
the
Lagrange formula for the case of operator matrices (cf. \cite{Na67, MM03, KS05, KS05a})
\ba{s13}
\Omega(\bu,\bv):=<{\bf L} \bu,\bv >-<\bu,{\bf L}^\dagger \bv>=\fv^* \cL  \fu,
\ea
with the adjoint differential expression \cite{Na67,MM03}
\be{s12}
{\bf L}^\dagger
\bv:=\sum_{q=0}^m (-1)^{m-q} \partial_x^{m-q} \left(\bl_q^*
\bv\right),
\ee
the vector $\fv$
\be{s14} \fv^T:=\left(\bv^T(0),\bv^{(1)T}_x(0),\ldots,
\bv^{(m-1)T}_x(0),\bv^T(1),\bv^{(1)T}_x(1),\ldots,
\bv^{(m-1)T}_x(1)\right)\in \CC^{2mN}
\ee
and the block matrix $\cL:=(\bl_{ij})$
\be{s15}
\cL=\left(\begin{array}{cc}
  -\fL (0) &  0\\
    0 & \fL (1)
\end{array}\right),\quad \fL(x)=\left(\begin{array}{ccccc}
 \bl_{00} &\bl_{01}&\cdots &\bl_{0 m-2}&\bl_{0 m-1}\\
 \bl_{10} & \bl_{11}&\cdots & \bl_{1 m-2} & 0  \\
  \vdots & \vdots & \iddots & \vdots & \vdots \\
 \bl_{m-2 0} & \bl_{m-2 1}& \cdots & 0& 0\\
 \bl_{m-1 0} & 0& \cdots & 0& 0
\end{array}\right),
\ee
where the matrices $\bl_{ij}$ are
\ba{s17}
\bl_{ij}&:=&\sum_{k=i}^{m-1-j}(-1)^k M^k_{ij}\,
\partial_x^{k-i}
\bl_{m-1-j-k},\nn \\
M^k_{ij}&:=&\left\{
\begin{array}{ll}
  \frac{k!}{(k-i)!i!}, & i+j\le m-1 \quad \cap \quad k\ge i\ge 0\\
   &  \\
  0, & i+j>m-1 \quad \cup \quad k<i.
\end{array}
\right.
\ea

Extend the original matrix $\fU$ (cf. \rf{s7}) to a square matrix $\cU$, which is
made non-degenerate in a neighborhood of the point ${\bf p}={\bf p}_0$ and the eigenvalue $\lambda=\lambda_0$
by an appropriate choice of the auxiliary matrices $\widetilde{ \fA}(\lambda,{\bf p})$ and
$\widetilde{ \fB}(\lambda,{\bf p})$
\be{s22}
\fU=[\fA,\fB] \hookrightarrow
\cU:=\left(\begin{array}{cc}
 \fA &\fB\\
 \widetilde{ \fA} &\widetilde{ \fB}\\
\end{array}\right)\in \CC^{2mN\times 2mN},\quad \widetilde{\fU}:=[\widetilde{ \fA},\widetilde{
\fB}],\quad \det (\cU)\neq 0.
\ee
Similar transformation for the adjoint boundary conditions $\fV \fv=[\fC,\fD]\fv=0$ yields
\be{s23}
\fV:=[\fC,\fD]\hookrightarrow \cV:=\left(\begin{array}{cc}
 \fC &\fD\\
 \widetilde{ \fC} &\widetilde{ \fD}\\
\end{array}\right)\in \CC^{2mN\times 2mN},\quad \widetilde{\fV}:=[\widetilde{ \fC},\widetilde{
\fD}], \quad \det (\cV)\neq 0.
\ee
Then, the form $\Omega(\bu,\bv)$ in \rf{s13} can be represented as \cite{Na67}
\be{s26}
\Omega(\bu,\bv)=(\fV \fv)^* \widetilde{\fU}\fu-(\widetilde{\fV}\fv)^*\fU\fu,
\ee
so that without loss in generality we can assume \cite{KS05, KS05a}
\be{s28}
\cL=\fV^*
\widetilde{\fU}-\widetilde{\fV}^*\fU.
\ee
Differentiating the equation \rf{s28} we find
\be{s29}
\partial^r_{\lambda}\cL=\sum_{k=0}^{r}\left(r \atop
k\right) \left[ \left(\partial_{\bar\lambda}^{r-k}\fV
\right)^*\partial_{\lambda}^{k}\widetilde{\fU}-
\left(\partial_{\bar\lambda}^{r-k}\widetilde{\fV}
\right)^*\partial_{\lambda}^{k}{\fU} \right].
\ee
Hence, we obtain the formula for calculation of the matrix $\fV$ of the adjoint
boundary conditions and the auxiliary matrix
$\widetilde{ \fV}$
\be{s31}
\left[\begin{array}{c}
 -\widetilde{ \fV}  \\
 \fV
 \\
\end{array}\right]^*=\cL \cU^{-1}=\left(\begin{array}{cc}
  -\fL (0) &  0\\
    0 & \fL (1)
\end{array}\right)\left(\begin{array}{cc}
 \fA &\fB\\
 \widetilde{ \fA} &\widetilde{ \fB}\\
\end{array}\right)^{-1},
\ee
which exactly reproduces and extends the corresponding result of \cite{KS05, KS05a}.

\section{Perturbation of eigenvalues}

Assume that in the neighborhood of the point ${\bf p}={\bf p}_0$
the spectrum of the boundary eigenvalue problem \rf{s1} is discrete. Denote
${\bf L}_0{=}{\bf L}(\lambda_0,{\bf p}_0)$ and ${\fU}_0{=}{\fU}(\lambda_0,{\bf p}_0)$.
Let us consider a smooth perturbation of parameters in the form ${\bf p}={\bf p}(\varepsilon)$
where ${\bf p}(0)={\bf p}_0$ and $\varepsilon$ is a small real number.
Then, as in the case of analytic matrix functions \cite{Ki05},
the Taylor decomposition of the differential operator matrix ${\bf L}(\lambda, {\bf p}(\varepsilon))$ and
the matrix of the boundary conditions ${\fU}(\lambda, {\bf p}(\varepsilon))$ are \cite{KS02, KS04, KS05, KS05a}
\be{p2}
{\bf L}\left(\lambda, {\bf p}(\varepsilon)\right) = \sum_{r,s=0}^{\infty}
\frac{(\lambda-\lambda_0)^r}{r!} \varepsilon^s\, {\bf L}_{rs}, \quad
{\fU}(\lambda, \varepsilon) =
\sum_{r,s=0}^{\infty}\frac{(\lambda-\lambda_0)^r}{r!}\varepsilon^s {\fU}_{rs},
\ee
with
\ba{p3}
{\bf L}_{00}&=&{\bf L}_0,\quad {\bf L}_{r0}={\partial^{r}_{\lambda}{\bf L}},\quad
{\bf L}_{r1}=
\sum_{j=1}^n\dot{p_j}\,
{\partial^{r}_{\lambda}\partial_{p_j} {\bf L}},\nn \\
\fU_{00}&=&\fU_0,\quad {\fU}_{r0}=
{\partial^{r}_{\lambda}{\fU}},\quad {\fU}_{r1}=
\sum_{j=1}^n\dot{p_j}\,
{\partial^{r}_{\lambda}\partial_{p_j} {\fU}},
\ea
where dot denotes differentiation with respect to $\varepsilon$ at $\varepsilon=0$
and partial derivatives are evaluated at ${\bf p}={\bf p}_0$, $\lambda=\lambda_0$.
Our aim is to derive explicit expressions for the leading terms in the expansions for multiple-semisimple and non-derogatory
eigenvalues and for the corresponding eigenvectors.

\subsection{Semi-simple eigenvalue}

Let at the point ${\bf p}={\bf p}_0$ the spectrum contain a semi-simple $\mu$-fold eigenvalue $\lambda_0$
with $\mu$ linearly-independent eigenvectors $\bu_0(x)$, $\bu_1(x)$,
$\ldots$, $\bu_{\mu-1}(x)$. Then, the perturbed eigenvalue $\lambda(\varepsilon)$ and the eigenvector ${\bf u}(\varepsilon)$
are represented as Taylor series in $\varepsilon$
\cite{VL60,KS04,SMK05,KMS05,GK06}
\be{p6}
\lambda=\lambda_0+\varepsilon \lambda_1 +\varepsilon^2
\lambda_2+\ldots, \quad \bu=\bb_0+\varepsilon \bb_1 +\varepsilon^2
\bb_2 +\ldots.
\ee

Substituting expansions \rf{p2} and \rf{p6} into \rf{s1} and collecting
the terms with the same powers of $\varepsilon$ we derive the boundary value problems
\be{p7}
{\bf L}_{0} {\bf b}_0=0,~~ {\fU}_{0} {\fb}_0 =0,
\ee
\be{p8}
{\bf L}_{0} {\bf b}_1 + (\lambda_1 {\bf L}_{10} + {\bf L}_{01}) {\bf b}_0=0,~~
{\fU}_{0} {\fb}_1 + (\lambda_1 {\fU}_{10} + {\fU}_{01}) {\fb}_0=0,
\ee
Scalar product of \rf{p8} with the eigenvectors $\bv_j$, $j=0,1,\ldots,\mu-1$ of the adjoint boundary
eigenvalue problem
\be{p10}
{\bf L}^\dagger_0 \bv=0,~~{\fV}_0{\fv}=0
\ee
yields $\mu$ equations
\be{p11}
<{\bf L}_0 \bb_1, \bv_j>=-\,<{\bf L}_{01} \bb_0, \bv_j>-\,\lambda_1
<{\bf L}_{10} \bb_0, \bv_j>.
\ee
With the use of the Lagrange formula \rf{s13}, \rf{s28} and the boundary conditions \rf{p8}
the left hand side of \rf{p11} takes the form
\be{p13}
<{\bf L}_0 \bb_1, \bv_j>
=(\widetilde{\fV}_0 {\fv}_j)^*({\fU}_{01}{\fb}_0+\lambda_1{\fU}_{10}\,{\fb}_0).
\ee
Together \rf{p11} and \rf{p13} result in the equations
\be{p14}
\lambda_1\left(<{\bf L}_{10}\, \bb_0,
\bv_j>+\,(\widetilde{\fV}_0
{\fv}_j)^*{\fU}_{10}\,{\fb}_0\right)=- <{\bf L}_{01} \bb_0,
\bv_j>-\,\,(\widetilde{\fV}_0 {\fv}_j)^*{\fU}_{01}{\fb}_0.
\ee
Assuming in the equations \rf{p14} the vector $\bb_0(x)$ as a linear combination
\be{p15}
\bb_0(x)=c_0 \bu_0(x) + c_1 \bu_1(x) + \ldots + c_{\mu-1} \bu_{\mu-1}(x),
\ee
and taking into account that
\be{p16}
{\fb}_0=c_0 {\fu}_0 + c_1 {\fu}_1 + \ldots + c_{\mu-1}
{\fu}_{\mu-1},
\ee
where ${\bf c}^T=(c_0,c_1,\ldots,c_{\mu-1})$, we arrive at the matrix eigenvalue problem (cf. \cite{LMZ03})
\be{p17}
-{\bf F}{\bf c}=\lambda_1 {\bf G}{\bf c}.
\ee
The entries of the $\mu \times \mu$ matrices $\bf F$ and $\bf G$ are defined by the expressions
\be{p18}
F_{ij}=<{\bf L}_{01} \bu_j, \bv_i>+\,\,
{\fv}_i^*\widetilde{\fV}_0^*{\fU}_{01}{\fu}_j,\quad
G_{ij}=\,<{\bf L}_{10} \bu_j, \bv_i> +\,\,
{\fv}_i^*\widetilde{\fV}_0^*\,{\fU}_{10}\,{\fu}_j.
\ee
Therefore, in the first approximation the splitting of the semi-simple eigenvalue due to variation
of parameters ${\bf p}(\varepsilon)$ is $\lambda=\lambda_0+\varepsilon \lambda_1 +o(\varepsilon)$,
where the coefficients $\lambda_1$ are generically $\mu$ distinct roots of the $\mu$-th order polynomial
\be{p19}
\det({\bf F}+\lambda_1{\bf G})=0.
\ee
For $\mu=1$ the formulas \rf{p18} and \rf{p19} describe perturbation of a simple eigenvalue
\be{p20} \lambda= \lambda_0- \varepsilon\frac{<{\bf L}_{01} \bu_0, \bv_0>+\,\,
{\fv}_0^*\widetilde{\fV}_0^*{\fU}_{01}{\fu}_0}{< {\bf L}_{10}
\bu_0, \bv_0> +\,\,
{\fv}_0^*\widetilde{\fV}_0^*\,{\fU}_{10}\,{\fu}_0}+o(\varepsilon).
\ee
The formulas \rf{p18},\rf{p19}, and \rf{p20} generalize the corresponding results of the works \cite{KS04} to the case
of the multiparameter non-self-adjoint boundary eigenvalue problems for operator matrices.

\subsection{Non-derogatory eigenvalue}

Let at the point ${\bf p}={\bf p}_0$ the spectrum contain a $\mu$-fold eigenvalue $\lambda_0$
with the Keldysh chain of length $\mu$, consisting of the
eigenvector $\bu_0(x)$ and the associated vectors $\bu_1(x),
{\ldots}, \bu_{\mu-1}(x)$. The vectors of the Keldysh chain solve
the following boundary value problems \cite{Na67,MM03}
\be{p21}
{\bf L}_0 \bu_0=0, \quad{\fU}_0{\fu}_0=0,\\
\ee
\be{p22} {\bf L}_0 \bu_j=- \sum_{r=1}^{j} \frac{1}{r!}
\partial^{r}_{\lambda}{\bf L}\bu_{j-r}, \quad
{\fU}_0{\fu}_j=-\sum_{r=1}^{j}
\frac{1}{r!}\partial^{r}_{\lambda}{\fU}{\fu}_{j-r}.
\ee

Consider vector-functions $\bv_0(x)$, $\bv_1(x)$, $\ldots$, $\bv_{\mu-1}(x)$.
Let us take scalar product of the differential equation \rf{p21} and
the vector-function $\bv_{\mu-1}(x)$. For each $j=1, \ldots, \mu-2$
we take the scalar product of the equation \rf{p22} and the
vector-function $\bv_{\mu-1-j}(x)$. Summation of the results yields the expression
\be{p23}
\sum_{j=0}^{\mu-1}\sum_{r=0}^{j} \frac{1}{r!}
<\partial^{r}_{\lambda}{\bf L}\bu_{j-r},\bv_{\mu-1-j}> = 0 \ee
Applying the Lagrange identity \rf{s13}, \rf{s28} and taking into account relation \rf{s29}, we transform \rf{p23}
to the form
\ba{p24}
\sum_{j=0}^{\mu-1}<\bu_{\mu-1-j},\sum_{r=0}^{j}\frac{1}{r!}\partial^{r}_{\bar
\lambda}{\bf L}^{\dagger}\bv_{j-r}>&+&\nn\\
\sum_{k=0}^{\mu-1}\sum_{j=0}^{\mu-1-k}\left[\sum_{r=0}^{j}
\left(\frac{1}{r!}\partial^{r}_{\bar\lambda}\fV\fv_{j-r}\right)^*\right]
\frac{\partial_{\lambda}^{k}\widetilde{\fU}}{k!}\fu_{\mu-1-j-k}&=&0.
\ea
Equation \rf{p24} is satisfied in case when the vector-functions $\bv_0(x)$, $\bv_1(x)$, $\ldots$, $\bv_{\mu-1}(x)$
originate the Keldysh chain of the adjoint boundary value problem, corresponding to the $\mu$-fold eigenvalue $\bar
\lambda_0$ \cite{KS05,KS05a}
\be{p24a}
{\bf L}_0^{\dagger}\bv_0=0,~~ {\fV}_0{\fv}_0=0,
\ee
\be{p25a}
{\bf L}_0^{\dagger}\bv_j=-\sum_{r=1}^{j} \frac{1}{r !}
\partial^{r}_{\bar \lambda}{{\bf L}^{\dagger}}\bv_{j-r},~~
{\fV}_0 {\fv}_j=-\sum_{r=1}^{j} \frac{1}{r !}
\partial^{r}_{\bar \lambda}{\fV}{\fv}_{j-r}.
\ee

Taking the scalar product of equation \rf{p22} and the vector ${\bf v}_0$ and employing the expressions \rf{s13}, \rf{s28}
we arrive at the orthogonality conditions
\be{p25} \sum_{r=1}^{j}\frac{1}{r!}\left[<\partial_{\lambda}^{r}{\bf L}
\bu_{j-r}, \bv_0>+\,\, {\fv}_0^*\widetilde{\fV}_0^*
\partial_{\lambda}^{r}{\fU}{\fu}_{j-r}\right]=0,\quad j=1, \ldots,
\mu-1.
\ee

Substituting into equations \rf{s1} the Newton-Puiseux series for the perturbed
eigenvalue $\lambda(\varepsilon)$ and eigenvector ${\bf u}(\varepsilon)$ \cite{K66,Ba84,KS05,KS05a}
\be{p26}
\lambda = \lambda_0 + \lambda_1 \varepsilon^{1/\mu} +\ldots, \quad
\bu= \bw_0 + \bw_1 \varepsilon^{1/\mu} +\ldots,
\ee
where $\bw_0=\bu_0$, taking into account expansions \rf{p2} and \rf{p26}
and collecting terms with the same powers of $\varepsilon$, yields
$\mu-1$ boundary value problems
serving for determining the functions ${\bf w}_r$, $r=1,2,\ldots, \mu-1$
\be{p28}
{\bf L}_0 \bw_r = -
\sum_{j=0}^{r-1} \left( \sum_{\sigma=1}^{r-j} \frac{1}{\sigma!}
{\bf L}_{\sigma0} \sum_{|\alpha|_{\sigma}=r-j}
\lambda_{\alpha_1} \ldots \lambda_{\alpha_{\sigma}} \right)\bw_j ,
\ee
\be{p29} {\fU}_0 {\fw}_r= - \sum_{j=0}^{r-1} \sum_{\sigma=1}^{r-j}
\left( \sum_{|\alpha|_{\sigma}=r-j} \lambda_{\alpha_1} \ldots
\lambda_{\alpha_{\sigma}} \right)\frac{1}{\sigma!}
{\fU}_{\sigma0}{\fw}_{\!\!j},
\ee
where $|\alpha|_{\sigma}=\alpha_1{+\ldots+}\alpha_{\sigma}$ and $\alpha_1$, $\ldots$,
$\alpha_{\mu-1}$ are positive integers. The vector-function $\bw_{\mu}(x)$ is a solution
of the following boundary value problem
\be{p31} {\bf L}_0 \bw_{\mu} =-{\bf L}_{01} \bw_0 - \sum_{j=0}^{\mu-1} \left(
\sum_{\sigma=1}^{\mu-j} \frac{1}{\sigma!}
{\bf L}_{\sigma0} \sum_{|\alpha|_{\sigma}=\mu-j}
\lambda_{\alpha_1} \ldots \lambda_{\alpha_{\sigma}} \right)\bw_j,
\ee
\be{p32} {\fU}_0  {\fw}_{\mu} =- {\fU}_{01} {\fw}_0  -
\sum_{j=0}^{\mu-1}
 \sum_{\sigma=1}^{\mu-j} \left(\sum_{|\alpha|_{\sigma}=\mu-j}
\lambda_{\alpha_1} \ldots \lambda_{\alpha_{\sigma}}
\right)\frac{1}{\sigma!}
{\fU}_{\sigma0} {\fw}_j.
\ee
Comparing equations \rf{p31} and \rf{p32} with the expressions \rf{p22}
we find the first $\mu-1$ functions $\bw_r$ in the expansions \rf{p26}
\be{p33}
\bw_r = \sum_{j=1}^{r} \bu_j
\sum_{|\alpha|_j=r} \lambda_{\alpha_1} \ldots \lambda_{\alpha_j}.
\ee
With the vectors \rf{p33} we transform the equations \rf{p31} and \rf{p32} into
\be{p34}
{\bf L}_0 \bw_{\mu} =-{\bf L}_{01} \bu_0- \lambda_1^{\mu} \sum_{r=1}^{\mu}
\frac{1}{r!}
\partial^{r}_{\lambda}{\bf L} \bu_{\mu-r}+
\sum_{j=1}^{\mu-1} {\bf L}_0\bu_j \sum_{|\alpha|_j=\mu} \lambda_{\alpha_1}
\ldots \lambda_{\alpha_j},
\ee
\be{p35} {\fU}_0 {\fw}_{\mu} =-{\fU}_1
{\fu}_0- \lambda_1^{\mu} \sum_{r=1}^{\mu} \frac{1}{r!}
\partial^{r}_{\lambda}{\fU} {\fu}_{\mu-r}+
\sum_{j=1}^{\mu-1} {\fU}_0{\fu}_j \sum_{|\alpha|_j=\mu}
\lambda_{\alpha_1} \ldots \lambda_{\alpha_j}.
\ee
Applying the expression following from the Lagrange formula
\ba{p36} <{\bf L}_0 \bw_{\mu}, \bv_0> &=& {\fv}_0^*\widetilde{\fV}_0^*
{\fU}_{01} {\fu}_0 +\lambda_1^{\mu} \sum_{r=1}^{\mu}
\frac{1}{r!}{\fv}_0^*\widetilde{\fV}_0^*
{\fU}_{r0} {\fu}_{\mu-r}\nn \\ &-&
\sum_{j=1}^{\mu-1} {\fv}_0^* \widetilde{\fV}_0^* {\fU}_0{\fu}_j
\sum_{|\alpha|_j=\mu} \lambda_{\alpha_1} \ldots \lambda_{\alpha_j},
\ea
and taking into account the equations for the adjoint Keldysh chain \rf{p24a} and
\rf{p25a} yields the coefficient $\lambda_1$ in \rf{p26}. Hence, the splitting of the $\mu$-fold
non-derogatory eigenvalue $\lambda_0$ due to perturbation of the parameters ${\bf p}={\bf p}(\varepsilon)$ is described by
the following expression, generalizing the results of the works \cite{KS02,KS04,KS05,KS05a}
\be{p37}
\lambda = \lambda_0 + \sqrt[\mu]{-\varepsilon \frac{< {\bf L}_{01} \bu_0, \bv_0>+\,\,
{\fv}_0^*\widetilde{\fV}_0^* {\fU}_{01}
{\fu}_0}{\sum_{r=1}^{\mu}\frac{1}{r!}
(<{\bf L}_{r0}\bu_{\mu{-}r},\bv_0>+\,\,
{\fv}_0^*\widetilde{\fV}_0^*{\fU}_{r0}{\fu}_{\mu-r})}}+o(\varepsilon^{\frac{1}{\mu}}).
\ee
For $\mu=1$ equation \rf{p37} is reduced to the equation \rf{p20} for a simple
eigenvalue.

\section{Example 1: A rotating circular string}

\begin{figure}
    \begin{center}
    \includegraphics[angle=0, width=0.8
    \textwidth]{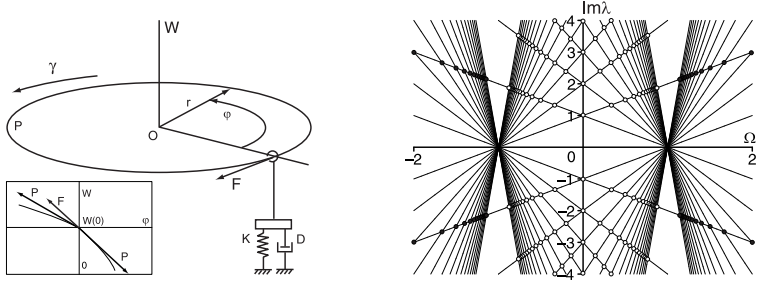}
    \end{center}
    \caption{A rotating circular string and its spectral mesh (only 30 modes are shown).}
    \label{fig1}
    \end{figure}

Consider a circular string of displacement
$W(\varphi, \tau)$, radius $r$, and mass per unit length $\rho$ that rotates with the speed $\gamma$ and
passes at $\varphi=0$ through a massless eyelet
generating a constant frictional follower force $F$ on the
string, as shown in Fig.~\ref{fig1}.
The circumferential tension $P$ in the string is assumed to be constant;
the stiffness of the spring supporting the eyelet is $K$ and the damping coefficient of the viscous damper is $D$.
Introducing the non-dimensional variables and parameters
\be{st1}
t=\frac{\tau}{r}\sqrt{\frac{P}{\rho}},\quad w=\frac{W}{r},\quad \Omega=\gamma r\sqrt{\frac{\rho}{P}},
\quad k=\frac{Kr}{P},\quad \mu=\frac{F}{P}, \quad d=\frac{D}{\sqrt{\rho P}},
\ee
and assuming  $w(\varphi, t)=u(\varphi)\exp(\lambda t)$ we arrive at
the non-self-adjoint boundary eigenvalue problem for a scalar $(N=1)$ differential operator \cite{YH95}
\be{st4}
Lu=\lambda^2 u+ 2\Omega\lambda u'-(1-\Omega^2)u''=0,
\ee
\be{st5}
u(0)-u(2\pi)=0,\quad u'(0)-u'(2\pi)=\frac{\lambda d+k}{1-\Omega^2}u(0)+\frac{\mu}{1-\Omega^2}u'(0),
\ee
where prime denotes differentiation with respect to $\varphi$.
Parameters
$\Omega$, $d$, $k$, and $\mu$ express the speed of rotation, and damping, stiffness, and
friction coefficients.

For the unconstrained rotating string with $d=0$, $k=0$, and $\mu=0$
the eigenfunctions $v$ and $u$ of the adjoint problems, corresponding to purely imaginary
eigenvalue $\lambda$ and $\bar \lambda$, coincide. With
$u$ $=$ $C_1 \exp\left(\varphi\lambda/(1-\Omega) \right)$ $+$ $ C_2 \exp\left(-\varphi\lambda/(1+\Omega)\right)$
assumed as a solution of \rf{st4}, the characteristic equation follows from \rf{st5}
\be{st8}
8\lambda\sin\frac{\pi\lambda}{i(1-\Omega)}\sin\frac{\pi\lambda}{i(1+\Omega)}
\frac{{\rm e}^\frac{-2\pi\lambda\Omega}{\Omega^2-1}
}{\Omega^2-1}=0.
\ee

The eigenvalues $\lambda_n^{\pm}=in(1\pm \Omega)$ with the eigenfunctions
$u_n^{\pm}=\cos(n \varphi)\mp i\sin(n \varphi)$, $n\in \mathbb{Z}$, found from equation \rf{st8}
form the spectral mesh in the plane $(\Omega,{\rm Im}\lambda)$, Fig.~\ref{fig1}.
The lines
$\lambda_n^{\varepsilon}=in(1+\varepsilon\Omega)$ and $\lambda_m^{\delta}=im(1+\delta\Omega)$,
where $\varepsilon,\delta=\pm1$, intersect each other at the node $(\Omega_{mn}^{\varepsilon \delta},\lambda_{mn}^{\varepsilon\delta})$ with
\be{st11}
\Omega_{mn}^{\varepsilon\delta}=\frac{n-m}{m\delta-n\varepsilon},\quad
\lambda_{mn}^{\varepsilon\delta}=\frac{inm(\delta-\varepsilon)}{m\delta-n\varepsilon},
\ee
where the double eigenvalue $\lambda_{mn}^{\varepsilon\delta}$ has two linearly independent eigenfunctions
\be{st13}
u_n^{\varepsilon}=\cos(n \varphi)-{\varepsilon}i\sin(n \varphi),\quad u_m^{\delta}=\cos(m \varphi)-\delta i\sin(m\varphi).
\ee

Using the perturbation formulas for semi-simple eigenvalues \rf{p18} and \rf{p19} with
the eigenelements \rf{st11} and \rf{st13} we find an asymptotic expression
for the eigenvalues originated after the splitting
of the double eigenvalues at the nodes of the spectral mesh due to interaction of the rotating string with the
external loading system
\be{st19}
\lambda=\lambda_{nm}^{\varepsilon\delta}+i\frac{\varepsilon n +\delta m}{2}\Delta\Omega+
i\frac{n+m}{8\pi nm}(d\lambda_{nm}^{\varepsilon\delta}+k)+
\frac{\varepsilon+\delta}{8\pi}\mu\pm \sqrt{c},
\ee
with $\Delta\Omega=\Omega-\Omega_{nm}^{\varepsilon\delta}$ and
\ba{st20}
c&=&\left(i\frac{\varepsilon n - \delta m}{2}\Delta \Omega+
i\frac{m-n}{8\pi mn}(d\lambda_{nm}^{\varepsilon\delta}+k)+
\frac{\varepsilon-\delta}{8\pi}\mu \right)^2\nn\\
&-&
\frac{(d\lambda_{nm}^{\varepsilon\delta}+k-i\varepsilon n\mu)
(d\lambda_{nm}^{\varepsilon\delta}+k-i\delta m\mu)}{16\pi^2nm}.
\ea

    \begin{figure}
    \begin{center}
    \includegraphics[angle=0, width=0.8
    \textwidth]{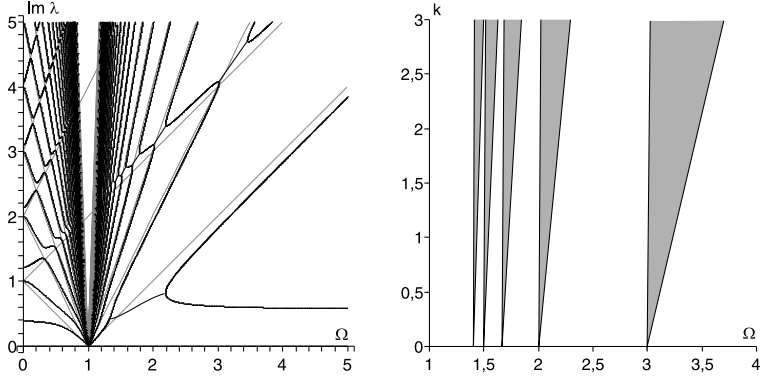}
    \end{center}
    \caption{Deformation of the spectral mesh of the rotating string
    interacting with the external spring with $k=2$ (left);
    approximation to the corresponding tongues of oscillatory instability (right).}
    \label{fig2}
    \end{figure}

Due to action of gyroscopic forces and an external spring double eigenvalues $\lambda_{nm}^{\varepsilon\delta}$
split in the subcritical region $|\Omega|<1$
($\varepsilon<0$, $\delta>0$ and $m>n>0$) as
\be{st21}
\lambda=\lambda_{nm}^{\varepsilon\delta}+
i\frac{m-n}{2}\Delta\Omega+i\frac{n+m}{8\pi nm}k\pm i\sqrt{\frac{k^2}{16\pi^2nm}+
\left(\frac{m-n}{8\pi mn}k-\frac{m+n}{2}\Delta\Omega\right)^2},
\ee
while in the supercritical region $|\Omega|>1$ ($\varepsilon<0$, $\delta>0$ and $m>0$, $n<0$)
\be{st22}
\lambda=\lambda_{nm}^{\varepsilon\delta}+
i\frac{m+|n|}{2}\Delta\Omega+i\frac{|n|-m}{8\pi |n|m}k\pm
\sqrt{\frac{k^2}{16\pi^2|n|m}-\left(\frac{|n|-m}{2}\Delta\Omega-\frac{m+|n|}{8\pi m|n|}k\right)^2}.
\ee
Therefore, for $|\Omega|<1$ the spectral mesh collapses into separated curves demonstrating avoided crossings;
for $|\Omega|>1$ the eigenvalue branches overlap forming the bubbles of instability with eigenvalues having positive real parts,
see Fig.~\ref{fig2}. From \rf{st22} a linear approximation follows to the boundary of the
domains of supercritical flutter instability in the plane $(\Omega, k)$ (gray resonance tongues in Fig.~\ref{fig2})
\be{st23}
k=\frac{4\pi|n|m(|n|-m)}{(\sqrt{|n|}\pm \sqrt{m})^2}\left(\Omega-\frac{|n|+m}{|n|-m}\right).
\ee

    \begin{figure}
    \begin{center}
    \includegraphics[angle=0, width=0.8\textwidth]{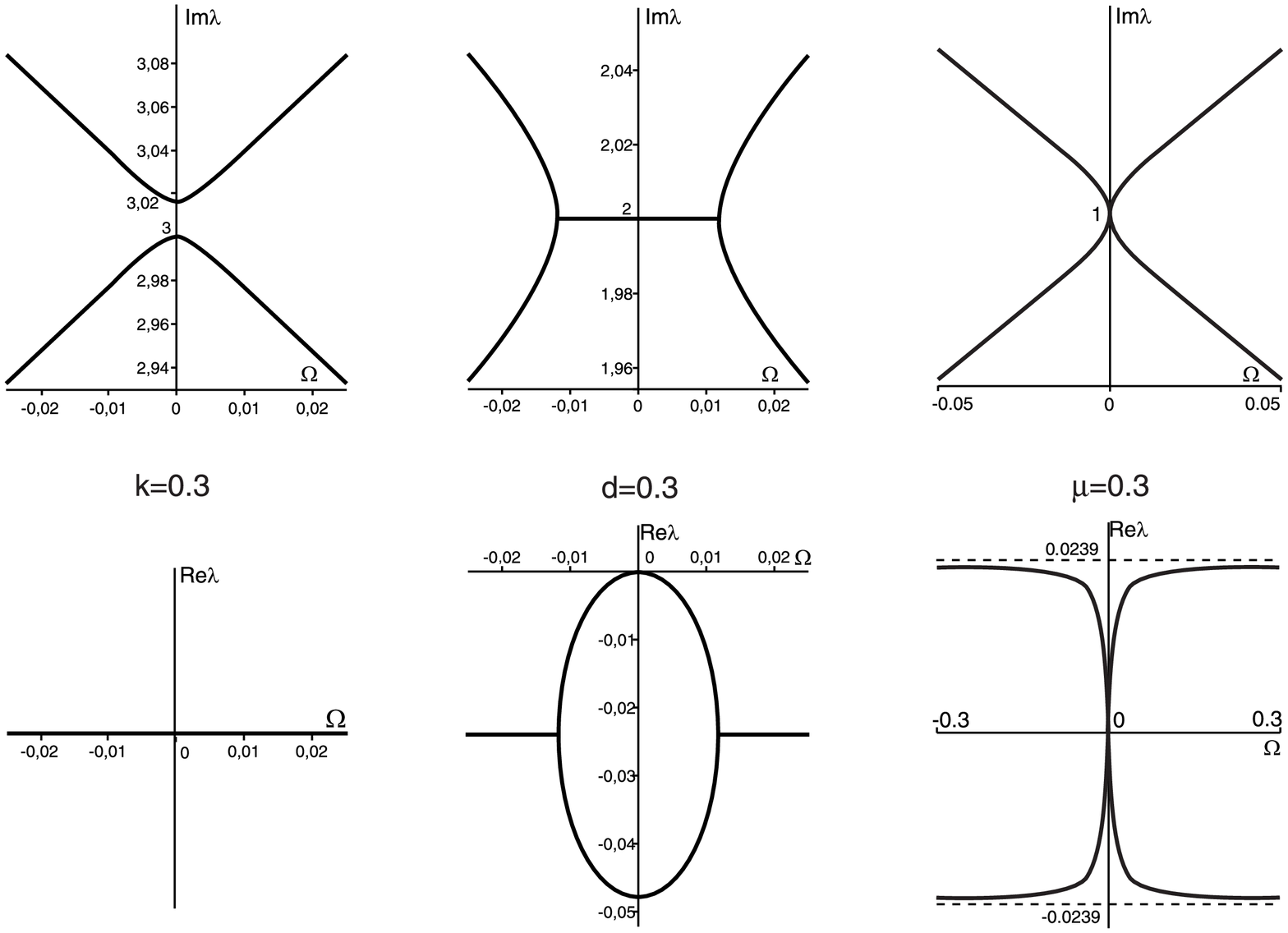}
    \end{center}
    \caption{Deformation of the spectral mesh of the string near the nodes $(0,3)$, $(0,2)$, and $(0,1)$.}
    \label{fig3}
    \end{figure}

At the nodes with $\Omega_{nm}^{\varepsilon\delta}=0$
the external damper creates a circle of complex eigenvalues being a latent source of subcritical flutter instability
responsible for the emission of sound in the squealing brake and the singing wine glass \cite{Ki08}
\be{st24}
\left({\rm Re}\lambda+\frac{d}{4\pi} \right)^2+n^2\Omega^2=\frac{d^2}{16\pi^2},\quad {\rm Im}\lambda=n,
\ee
\be{st25}
n^2\Omega^2-\left( {\rm Im}\lambda - n \right)^2=\frac{d^2}{16\pi^2},\quad
{\rm Re}\lambda=-\frac{d}{4\pi},
\ee
as shown in Fig.~\ref{fig3}.
Non-conservative perturbation yields eigenvalues with
\be{st26}
{\rm Im}\lambda=n\pm\frac{1}{2\pi}\sqrt{2\pi^2 n^2\Omega^2\pm\pi n\Omega \sqrt{4\pi^2n^2\Omega^2+\mu^2}},
\ee
\be{st27}
{\rm Re}\lambda=\pm\frac{1}{2\pi}\sqrt{-2\pi^2 n^2\Omega^2\pm\pi n\Omega\sqrt{4\pi^2n^2\Omega^2+\mu^2}},
\ee
so that
both the real and imaginary parts of the eigenvalue branches show a degenerate crossing,
touching at the node $(0,n)$, Fig.~\ref{fig3}.
Deformation patterns of the spectral mesh and first-order approximations of the instability tongues
obtained by the perturbation theory and shown in Fig.~\ref{fig2} and Fig.~\ref{fig3} are in a good
qualitative and quantitative agreement with the results of numerical calculations of \cite{YH95}.

\section{Example 2: MHD $\alpha^2$-dynamo}

Consider a non-self-adjoint boundary eigenvalue problem $(N=2)$
appearing in the theory of MHD $\alpha^2$-dynamo \cite{krause1,SGGX06,GK06,GKSS07}

\be{dm1}
{\bf L}{\bf u}:={\bf l}_0\partial_x^2 {\bu}
+
{\bf l}_1\partial_x\bu
+
{\bf l}_2{\bu}=0,\quad \fU\fu:=[\fA,\fB]\fu=0,
\ee
with the matrices of the differential expression
\be{dm2}
{\bf l}_0=\left(%
\begin{array}{cc}
  1 & 0 \\
  -\alpha(x) & 1 \\
\end{array}%
\right),\quad
{\bf l}_1=\partial_x {\bf l}_0,\quad
{\bf l}_2=
\left(%
\begin{array}{cc}
  -\frac{l(l+1)}{x^2}-\lambda & \alpha(x) \\
  \alpha(x)\frac{l(l+1)}{x^2} & -\frac{l(l+1)}{x^2} -\lambda\\
\end{array}%
\right),
\ee
and of the boundary conditions
\be{dm3}
\fA=\left(%
\begin{array}{cccc}
  1 & 0 & 0 & 0 \\
  0 & 1 & 0 & 0 \\
  0 & 0 & 0 & 0 \\
  0 & 0 & 0 & 0 \\
\end{array}%
\right),\quad
\fB=\left(%
\begin{array}{cccc}
  0 & 0 & 0 & 0 \\
  0 & 0 & 0 & 0 \\
  \beta l + 1 - \beta & 0 & \beta & 0 \\
  0 & 1 & 0 & 0 \\
\end{array}%
\right),
\ee
where it is assumed that $\alpha(x)=\alpha_0+\gamma\Delta\alpha(x)$ with $\int_0^1 \Delta\alpha(x) dx =0$.
For the fixed $\Delta\alpha(x)$ the differential expression
depends on the parameters $\alpha_0$ and $\gamma$, while $\beta$ interpolates between the idealistic ($\beta=0$)
and physically realistic ($\beta=1$) boundary conditions \cite{krause1,SGGX06,GK06,GK07}.

The matrix $\fV$ of the boundary conditions and auxiliary matrix $\widetilde{\fV}$
for the adjoint differential expression
$
{\bf L}^{\dagger}{\bf v}={\bf l}_0^*\partial^2_x {\bv}-{\bf l}_1^*\partial_x {\bv}+{\bf l}_2^* {\bv}
$
follow from the formula \rf{s31} where the $4\times4$ matrices $\widetilde{\fA}$, and $\widetilde{\fB}$ are chosen as
\be{dm5}
\widetilde{\fA}=\left(
                  \begin{array}{cc}
                    0 & {\bf I} \\
                    0 & 0 \\
                  \end{array}
                \right),\quad
\widetilde{\fB}=\left(
                  \begin{array}{cc}
                    0 & 0 \\
                    0 & {\bf I} \\
                  \end{array}
                \right).
\ee

\begin{figure}
    \begin{center}
    \includegraphics[angle=0, width=1.0\textwidth]{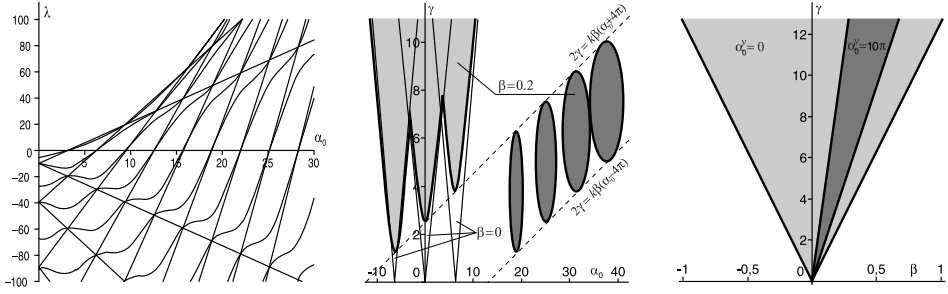}
    \end{center}
    \caption{$l=0$: Deformation of the spectral mesh for $\gamma=0$ and $\beta \in (0,1)$;
    approximation of the resonance tongues for
    $\Delta\alpha(x)=\cos(4\pi x)$ and
    $\lambda_0^{\nu}<0$ (white, light gray) or $\lambda_0^{\nu}>0$ (dark gray).}
    \label{fig4}
    \end{figure}

In our subsequent consideration we assume that $l=0$ and
interpret $\beta$ and $\gamma$ as perturbing parameters.
It is known \cite{GK06} that for $\beta=0$ and $\gamma=0$ the spectrum of
the unperturbed eigenvalue problem \rf{dm1}
forms the spectral mesh in the plane $(\alpha_0, \lambda)$, as shown in Fig.~\ref{fig4}.
The eigenelements of the spectral mesh are
\be{dm6}
\lambda_n^{\varepsilon}=-(\pi n)^2 + \varepsilon \alpha_0 \pi n,\quad
\lambda_m^{\delta}=-(\pi m)^2 + \delta \alpha_0 \pi m, \quad \varepsilon, \delta = \pm1,
\ee
\be{dm7}
{\bu^{\varepsilon}_n}=\left(
                        \begin{array}{c}
                          1 \\
                          \varepsilon \pi n \\
                        \end{array}
                      \right)\sin(n\pi r), \quad
{\bu^{\delta}_m}=\left(
                        \begin{array}{c}
                          1 \\
                          \delta \pi m \\
                        \end{array}
                      \right)\sin(m\pi r).
\ee
The branches \rf{dm6} intersect and
originate a double semi-simple eigenvalue with two linearly independent eigenvectors \rf{dm7}
at the node $(\alpha_0^{\nu},\lambda_0^{\nu})$, where \cite{GK06}
\be{dm8}
\lambda_0^{\nu}=\varepsilon\delta \pi^2 nm,\quad
\alpha_0^{\nu}=\varepsilon\pi n + \delta \pi m.
\ee
Taking into account that the components of the eigenfunctions of the adjoint
problems are related as $\bar v_2=u_1$ and $\bar v_1=u_2$, we find from equations \rf{p18} and \rf{p19} the asymptotic formula
for the perturbed eigenvalues, originating after the splitting of the double
semi-simple eigenvalues at the nodes of the spectral mesh
\ba{dm9}
\lambda&{=}&\lambda_0^{\nu}-\varepsilon\delta \pi^2 mn \beta+\frac{\pi}{2}(\delta m+\varepsilon n)\Delta\alpha_0 \\
&{\pm}& \frac{\pi}{2} \sqrt{\left((\delta m {-}\varepsilon n)\Delta\alpha_0 \right)^2+
4mn(\varepsilon \gamma\Delta\alpha {-} (-1)^{n+m} \pi n \beta)(\delta \gamma\Delta\alpha{-}(-1)^{n+m} \pi m \beta)},\nn
\ea
where
\be{dm10}
\Delta \alpha_0:=\alpha_0^{\nu}-\alpha_0, \quad
\Delta \alpha:=\int_0^1 \Delta \alpha (x) \cos((\varepsilon n - \delta m)\pi x)dx.
\ee
When $\gamma=0$ and $\Delta\alpha_0=0$, one of the two simple eigenvalues \rf{dm10} remains unshifted in all orders
of the perturbation theory with respect to the parameter $\beta$: $\lambda=\lambda_0^{\nu}$.
The sign of the first-order increment to another eigenvalue $\lambda=\lambda_0^{\nu}-2\lambda_0^{\nu}\beta$ depends
on the sign of $\lambda_0^{\nu}$,
which is directly determined by the Krein signature of the modes involved in the crossing \cite{GK06}.
This is in the qualitative and quantitative agreement with
the results of numerical calculations of \cite{GK07} shown in Fig.~\ref{fig4}.

Under variation of the parameter $\beta$ in the boundary conditions the eigenvalues remain real.
Additional parameter $\gamma$ is required for the creation of complex eigenvalues, which happens when
the radicand in \rf{dm9} becomes negative
\ba{dm12}
\left((\varepsilon n-\delta m)\Delta\alpha_0 \right)^2&+&
mn\left((\varepsilon+\delta)\gamma\Delta\alpha -(-1)^{n+m}(n+m)\beta\pi\right)^2\nn \\
&<&
mn\left((\varepsilon-\delta)\gamma\Delta\alpha -(-1)^{n-m}(n-m)\beta\pi\right)^2
\ea

Inequality \rf{dm12} defines the inner part of a cone in the space $(\alpha_0, \beta, \gamma)$.
The part of the cone corresponding to ${\rm Re}\lambda>0$ (oscillatory dynamo) is selected by the condition
\be{dm13}
\lambda_0^{\nu}-\varepsilon\delta \pi^2 mn \beta+\frac{\pi}{2}(\delta m+\varepsilon n)\Delta\alpha_0>0.
\ee

The conical zones develop according to the resonance selection rules discovered in \cite{GK06}. For example,
if $\Delta \alpha(x)=\cos(2\pi k x)$, then
\be{dm14}
\Delta \alpha=\int_0^1 \cos(2\pi k x) \cos((\varepsilon n - \delta m)\pi x)dx=
\left \{\begin{array}{l}
1/2,\quad 2k=\pm(\varepsilon n- \delta m) \\
0,\quad 2k\ne\pm(\varepsilon n- \delta m)
\end{array}
 \right.
\ee
There exist $2k-1$ cones for $\lambda_0^{\nu}<0$ $(\varepsilon\delta<0)$ and infinitely many cones for $\lambda_0^{\nu}>0$
$(\varepsilon\delta>0)$. Due to different inclinations of the cones, only
cross-sections of $2k-1$ cones with ${\rm Re}\lambda<0$ survive in the plane $\beta=0$. They are situated symmetrically with respect to the
$\gamma$-axis
\be{dm15}
\left(\Delta\alpha_0 \right)^2<
\frac{\gamma^2}{4}\left[ 1- \left(\frac{n-k}{k} \right)^2\right],\quad n=k, k+1,\ldots,2k-1.
\ee
For $k=2$ three resonant tongues $4\alpha_0^2<\gamma^2$ and $16\left(\alpha_0 \pm 2\pi \right)^2<3\gamma^2$
are shown white in Fig.~\ref{fig4}.
When $\beta \ne 0$ the tongues \rf{dm15}, corresponding to $\lambda_0^{\nu}<0$,
are deformed into the domains in the plane $(\alpha_0,\gamma)$ bounded by the hyperbolic curves
\be{dm16}
4k^2\left(\alpha_0 \pm \alpha_0^{\nu} \right)^2
+\left(( n - m)\gamma/2 \pm \pi (n^2+m^2) \beta\right)^2
<k^2\left(\gamma\pm 2\pi (n - m) \beta\right)^2.
\ee
The deformed resonant tongues are located at a distance $\gamma_0=\gamma(\Delta\alpha_0=0)$ from the $\alpha_0$-axis,
which for the tongues with $\alpha_0^{\nu}>0$ is greater $(\gamma_0=2\pi n \beta)$ than for those with
$\alpha_0^{\nu}<0$ $(\gamma_0=2\pi (2k-n) \beta)$.
In case when $k=2$ the approximation to the deformed principal resonant tongues
\be{dm17}
\gamma^2-4\alpha_0^2>16\pi^2\beta^2, \quad 16(\alpha_0\pm2\pi)^2+(\gamma\pm10\pi\beta)^2<4(\gamma\pm4\pi\beta)^2
\ee
is  shown light gray in Fig.~\ref{fig4}.

Cross-sections by the plane $\beta \ne 0$ of the cones, corresponding to $\lambda_0^{\nu}>0$,  have the form of the
ellipses, shown dark gray in Fig.~\ref{fig4}
\be{dm18}
4k^2\left(\alpha_0 \pm \alpha_0^{\nu} \right)^2+n(2k+n)\left(\gamma \pm 2(n+k)\pi\beta \right)^2<n(2k+n)4k^2\pi^2\beta^2,
\ee
where $n=1,2,\ldots$. The eigenvalues inside the ellipses have positive real parts, which corresponds to the excitation
of the \textit{oscillatory dynamo} regime. The ellipses belong to a corridor bounded by the lines $2\gamma=k\beta(\alpha_0\pm4\pi)$. Hence, the amplitude $\gamma$ of the resonant perturbation of the $\alpha$-profile
 $\gamma\Delta\alpha(x)$ is limited both from below and from above in agreement with the numerical findings
of \cite{SGGX06}.

When $\beta \rightarrow 0$, the ellipses shrink to the diabolical points $(\alpha_0^{\nu},0)$ in the plane
$(\alpha_0,\gamma)$. The reason for this effect is the inclination of the cone
\rf{dm18}. The rightmost picture of Fig.~\ref{fig4} shows a cross-section of
the inclined cone \rf{dm18} (dark gray) by the plane $\alpha_0=\alpha_0^{\nu}$ together with that of the cone \rf{dm16} (light gray). Obviously, variation of $\gamma$ (or, equivalently, of $\alpha$-profile) excites the complex eigenvalues only near the diabolical points with $\lambda_0^{\nu}<0$ in accordance with
\cite{GK06}, while the variation of $\beta$ does not produce the complex eigenvalues at all \cite{GK07}.
Nevertheless, the variation of $\beta$ together with $\gamma$ yields the complex eigenvalues
near the nodes of the spectral mesh with both $\lambda_0^{\nu}<0$ and $\lambda_0^{\nu}>0$.
We note that the analytical results are confirmed both qualitatively and quantitatively by the Galerkin-based
numerical simulations.

\section*{Conclusion}
A multiparameter perturbation theory for non-self-adjoint boundary eigenvalue problems for matrix differential operators
is developed in the form convenient for implementation in the computer algebra systems
for an automatic calculation of the adjoint boundary conditions and coefficients in the perturbation series for eigenvalues and eigenvectors.
The approach is aimed to the applications requiring frequent switches from one set of boundary conditions to another.
Two studies of the onset of instability in rotating continua under symmetry-breaking perturbations, demonstrate the efficiency
of the proposed approach.


\subsection*{Acknowledgment}
The author is grateful to Professor P. Hagedorn for useful discussions.
\end{document}